\documentclass[aps,prl,twocolumn,groupedaddress,showpacs,amsmath,amssymb]{revtex4}

\usepackage{graphicx}
\usepackage{dcolumn}
\usepackage{bm}

\begin{document}

\title{Scaling behavior of temperature-dependent thermopower in CeAu$_{2}$Si$_{2}$ under pressure}

\date{\today}
\author{Z. Ren$^{1}$}
\email{Zhi.Ren@unige.ch}
\author{G. W. Scheerer$^{1}$}
\author{G. Lapertot$^{2}$}
\author{D. Jaccard$^{1}$}
\affiliation{$^{1}$DQMP - University of Geneva, 24 Quai Ernest-Ansermet, 1211 Geneva 4, Switzerland}
\affiliation{$^{2}$SPSMS, UMR-E CEA/UJF-Grenoble 1, INAC, Grenoble, F-38054, France}

\begin{abstract}
We report a combined study of in-plane resistivity and thermopower of the pressure-induced heavy fermion superconductor CeAu$_{2}$Si$_{2}$ up to 27.8 GPa.
It is found that thermopower follows a scaling behavior in $T$/$T^{\rm *}$ almost up to the magnetic critical pressure $p_{\rm c}$ $\sim$ 22 GPa.
By comparing with resistivity results, we show that the magnitude and characteristic temperature dependence of thermopower in this pressure range are governed by the Kondo coupling and crystal-field splitting, respectively.
Below $p_{\rm c}$, the superconducting transition is preceded by a large negative thermopower minimum, suggesting a close relationship between the two phenomena.
Furthermore, thermopower of a variety of Ce-based Kondo-lattices with different crystal structures follows the same scaling relation up to $T$/$T^{\rm *}$$\sim$ 2.

\end{abstract}

\maketitle
\maketitle

\section{I. Introduction}

Thermopower ($S$) of Ce-based Kondo lattice systems (KLS) exhibits a variety of unusual features depending on the coupling strength ($J$) between
Ce-4$f$ and conduction electrons, which is of both fundamental and application interests \cite{Zlaticbook,Sakurai}. Particularly, as $J$ is enhanced by external pressure ($p$), typical features in $S(T)$ from Kondo to intermediate-valence (IV) regime can be observed even in a single compound,
examples of which include CeCu$_{2}$Si$_{2}$ \cite{CeCu2Si2}, CeAl$_{3}$ \cite{CeAl3}, CeCu$_{2}$Ge$_{2}$ \cite{CeCu2Ge2andCePd2Si2}, CePd$_{2}$Si$_{2}$ \cite{CeCu2Ge2andCePd2Si2}, CeCu$_{2}$ \cite{CeCu2}, CeRu$_{2}$Ge$_{2}$ \cite{CeRu2Ge2}.
At $p$ = 0, $S(T)$ of these materials displays a positive maximum at high $T$ but a negative minimum at low $T$.
The increase of $p$ tends to upshift the whole $S(T)$ curve. Once $J$ becomes large enough for Ce to be in an IV state,
one broad positive $S(T)$ maximum of the order of $k_{\rm B}$/$e$ (= 87 $\mu$V/K) persists, often with a shoulder at low $T$ \cite{CePd3}.
Theoretically, while at high and intermediate $T$ the $S$ behavior is understood as resulting from the interplay between Kondo and crystal field (CF) effects \cite{Coqblin,Zlatic1,Zlatic2},
the low-$T$ negative $S$ remains mysterious.

However, since in the aforementioned cases the long-range magnetic order (if any) disappears at $p$ $<$ 8 GPa \cite{CeCu2Ge2andCePd2Si2,CeCu2,CeRu2Ge2},
the way in which $S(T)$ evolves with $p$ remains largely unexplored for weak Kondo coupling (small $J$) regime.
In this context, it is worth noting that we recently found a giant overlap in $p$ between the magnetic and superconducting phases of CeAu$_{2}$Si$_{2}$, which contrasts radically with the observations made on its sister compounds CeCu$_{2}$$X$$_{2}$ ($X$ = Si, Ge) \cite{RenPRX}.
Strikingly, the magnetic order of CeAu$_{2}$Si$_{2}$ persists up to $p_{\rm c}$ $\sim$ 22 GPa \cite{RenPRX}, which, together with the small magnitude of $\mid$$S$$\mid$ ($\sim$2 $\mu$V/K) \cite{DidierCeCu2Ge2,CeAu2Si2TEP1,CeAu2Si2TEP2},
places the compound far from magnetic instabilities.
Thus, the high-$p$ $S$($T$) measurement of CeAu$_{2}$Si$_{2}$ allows not only to look for possible differences with the normal-state properties of CeCu$_{2}$$X$$_{2}$, but also offers a rare opportunity to study the $p$-evolution of the characteristic features in $S$($T$) starting from a small $J$.

In this paper, we present thermopower $S(T)$ and electrical resistivity $\rho(T)$ measurements on the very same CeAu$_{2}$Si$_{2}$ crystal at $p$ up to 27.8 GPa.
Thanks to its large $p_{\rm c}$, a remarkable scaling behavior is uncovered: the $S(T)$ data for $p$ $\leqslant$ 20.5 GPa, when normalized by a proper factor, collapse onto a single curve when plotted against $T$/$T^{\rm *}$, where $T^{\rm *}$ is the temperature at which the first sign change of $S$ occurs with decreasing $T$.
The comparison with the results of $\rho$ measurements shows that the normalization factor and $T^{\rm *}$ follow nearly the same $p$-dependence as the $-$ln$T$ slope of $\rho(T)$ and overall CF splitting $\Delta_{\rm CF}$, respectively.
In the $p$-range where superconductivity (SC) is enhanced with $p$, a large negative minimum in $S(T)$ develops below $T^{\rm *}$.
Further on, it is shown that the scaling relation is commonly applicable to Ce-based KLS provided that the coupling strength $J$ is small enough.
These results are discussed in regard to existing theories and their possible implication for Ce-based heavy fermion SC.

\section{II. Experimental}

We grow CeAu$_{2}$Si$_{2}$ crystals from Sn flux as described elsewhere \cite{RenPRX,OnukiCeAu2Si2}.
A sample cut from a crystal with low residual resistivity $\rho_{\rm 0}$ $\approx$ 2 $\mu$$\Omega$cm is used for in-plane $S$($T$) and $\rho$($T$) measurements, which are carried out in a Bridgman-type anvil pressure cell in the temperature range 1.3 $<$ $T$ $<$ 300 K with lead (Pb) as the $p$-gauge \cite{Bridgmananvil}.
A photograph of the high-$p$ chamber is shown in S1 of the Supplemental Material \cite{SM}.
Compared with the previous study \cite{RenPRX}, we succeed in further miniaturizing the setup by reducing the sample size, the $p$-cell thickness ($d$) and the cross section of the thermocouple (TC) wires.
Due to the rapid relaxation of the thermal gradient $\Delta$$T$ $\propto$ exp(-1/$d$), where $d$ $\approx$ 40 $\mu$m, special care must be taken to position the TC wires as close as possible to the heater.
Following Ref. \cite{CeRu2Ge2}, $S(T)$ of the TC wires are assumed to be $p$-independent, and we check that our results are free from their (possible) small variations under $p$.
Corrections due to the misplacement of the TC wires are introduced by examining the chamber after depressurization.
Within the experimental error of $\sim$ 20\%, the results presented here are in good agreement with those obtained in another cell for the overlapping $T$-range (1.3 $<$ $T$ $<$ 7 K) \cite{GernotCeAu2Si2}.

\begin{figure}
\includegraphics*[width=9cm]{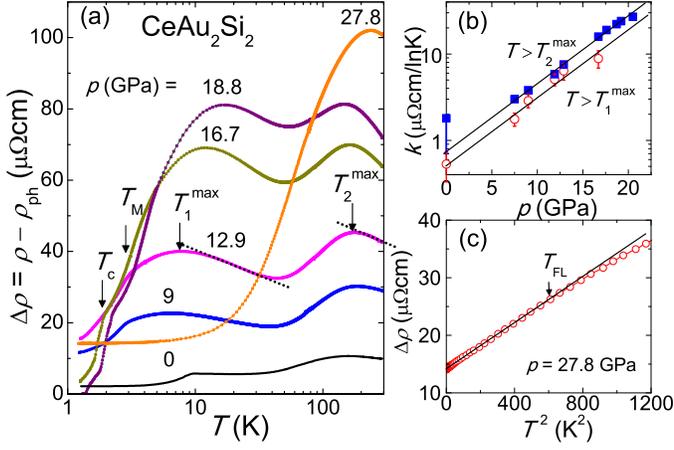}
\caption{(Color online)
(a) Logarithmic $T$-dependence of the non-phononic contribution ($\Delta$$\rho$) to in-plane resistivity of CeAu$_{2}$Si$_{2}$ for typical pressures.
The arrows indicate the two resistivity maxima at $T_{1}^{\rm max}$ and $T_{2}^{\rm max}$, respectively, for 12.9 GPa.
The two dashed lines evidence the --ln$T$ dependence of resistivity.
At 16.7 GPa, the magnetic and onset superconducting transitions are denoted by $T_{\rm M}$ and $T_{\rm c}$, respectively.
(b) The $-$ln$T$ slopes $k_{\rm 1}$ and $k_{\rm 2}$ above $T_{1}^{\rm max}$ and $T_{2}^{\rm max}$ as a function of $p$.
At $p$ = 0, the large error in $k_{\rm 2}$ is due to the uncertainty in $\rho_{\rm ph}$;
for the other pressures, the error is within the size of the symbol.
The solid lines are a guide to the eyes.
(c) $\Delta$$\rho$ at 27.8 GPa plotted a function of $T^{2}$. $T_{\rm FL}$ denotes the Fermi liquid temperature below which $\Delta$$\rho$ $\propto$ $T^{2}$ (solid line).
}
\label{fig1}
\end{figure}

\section{III. Results and Discussion}
Let us first briefly summarize the $\rho$($T$) results, which are essentially the same as in previous studies \cite{RenPRX,RenPRB}. Figure 1(a) shows the $T$-dependence of the non-phononic resistivity $\Delta$$\rho$ = $\rho$ $-$ $\rho_{\rm ph}$ for typical pressures, where $\rho_{\rm ph}$ is a phonon term derived from $\rho$($T$) of LaPd$_{2}$Si$_{2}$ \cite{LaPd2Si2} and assumed to be $p$-independent.
Note that such a $\rho_{\rm ph}$ is a better approximation at low $T$ than the linear one used previously \cite{RenPRX,RenPRB}, which overestimates the actual $\rho_{\rm ph}$ contribution.
At an intermediate $p$, say 12.9 GPa, and as observed in numerous KLS, $\Delta$$\rho$($T$) exhibits two maxima at $T_{\rm 1}^{\rm max}$ and $T_{\rm 2}^{\rm max}$. Moreover, above each maximum $\Delta$$\rho$($T$) $\propto$ $-$ln$T$, reflecting the incoherent Kondo scattering of the ground state and excited CF levels \cite{scattering1}.
The increase of $p$ has little effect on $T_{\rm 2}^{\rm max}$, but enhances both $T_{\rm 1}^{\rm max}$ and the $-$ln$T$ slope $k_{i}$ = $-$$d$($\Delta$$\rho$)/$d$(ln$T$) for $T$ $>$ $T_{i}^{\rm max}$ ($i$ = 1, 2) \cite{slope}.
Strikingly, as shown in Fig.1(b), $k_{\rm 1}$ and $k_{\rm 2}$ share almost the same $p$-exponential dependence up to 16.7 and 20.5 GPa \cite{Note1}, respectively, pointing to a $p$-independent ratio of $k_{\rm 2}$/$k_{\rm 1}$ $\approx$ 1.5.
It must be emphasized that the $p$-dependence of the $k_{\rm 1}$ and $k_{\rm 2}$ slopes is practically independent of the choice of $\rho_{\rm ph}$, but their ratio ($k_{\rm 2}$/$k_{\rm 1}$) can vary from 1.5 to 2.5 for different $\rho_{\rm ph}$ terms.
Well below $T_{\rm 1}^{\rm max}$, signature of $p$-induced SC is detected at 16.7 and 18.8 GPa in addition to magnetic ordering, though the transition is broadened likely due to $p$-gradient.
At the highest $p$ of 27.8 GPa, Fermi liquid (FL) behavior extends up to $T_{\rm FL}$ $\approx$ 25 K [Fig. 1(c)], and the Kondo temperature $T_{\rm K}$ $\approx$ 240 K, estimated from the $T_{\rm 1}^{\rm max}$ \cite{SunCeCu2Si2}, is $\sim$140 times the ambient-$p$ value (1.7 K) deduced from the neutron scattering experiments \cite{CeAu2Si2TK}.
\begin{figure}
\includegraphics*[width=8cm]{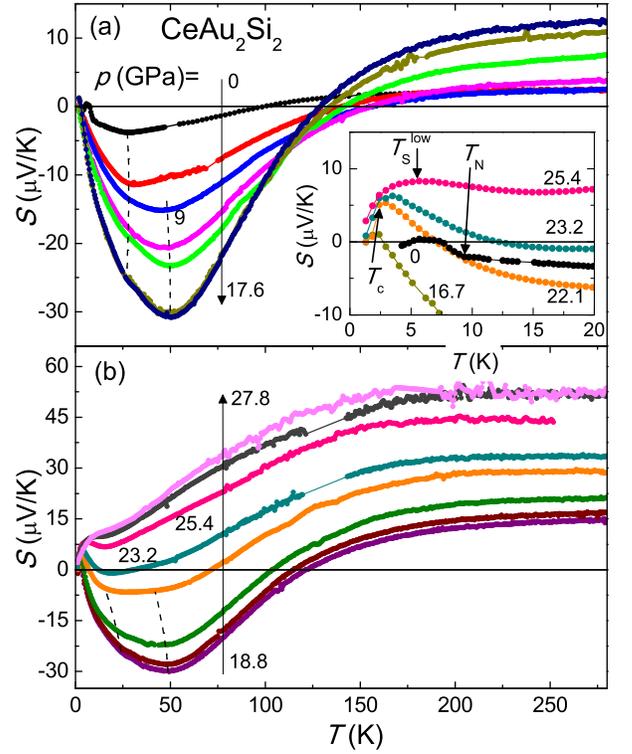}
\caption{(Color online)
In-plane $S$($T$) of CeAu$_{2}$Si$_{2}$ for (a) $p$ $\leqslant$ 17.6 GPa and (b) 18.8 $\leqslant$ $p$ $\leqslant$ 27.8 GPa.
The vertical arrows and dashed lines are a guide to the eyes.
The inset shows the low temperature data for typical pressures.
The temperatures corresponding to the superconducting transition and the maximum in thermopower are marked by arrows.
}
\label{fig2}
\end{figure}

We will now discuss the in-plane $S$($T$) data.
At $p$ = 0, $S(T)$ of CeAu$_{2}$Si$_{2}$ undergoes a sign change at $T^{\rm *}$ $\approx$ 100 K
and shows a minimum at $T_{\rm min}$ $\approx$ 25 K.
This nonmonotonic $T$-dependence is in stark contrast to that found in polycrystalline non-magnetic LaAu$_{2}$Si$_{2}$ \cite{CeAu2Si2TEP2},
suggesting that the dominant contribution to $S(T)$ already stems from weak Kondo scattering.
As can be seen in Fig. 2(a) and (b), the evolution of $S$($T$) at low $T$ is qualitatively different at $p$ below and above 17.6 GPa.
With increasing $p$ below 17.6 GPa, it is observed for the first time that the magnitude of the negative $S(T)$ is boosted from small to giant values typical of KLS.
Concomitantly, $T_{\rm min}$ increases to $\sim$50 K for $p$ $\geqslant$ 9 GPa.
On closer examination, a bump at $\sim$25 K is still discernible, suggesting that there are actually two superimposed negative contributions (see dashed lines in Fig. 2).
At 17.6 GPa, the $S_{\rm min}$ value $\sim$$-$30 $\mu$V/K  is close to that of CeCu$_{2}$Si$_{2}$ at $p$ = 0 \cite{CeCu2Si2,SunCeCu2Si2},
an expected result given that at this $p$ the volume of CeAu$_{2}$Si$_{2}$ is reduced to that of CeCu$_{2}$Si$_{2}$\cite{RenPRX}.
Since the overall $S$($T$) behavior of CeCu$_{2}$Si$_{2}$ is very similar along the $ab$ plane and $c$-axis with nearly isotropic $S_{\rm min}$ \cite{unpublished},
a weak anisotropy in $S$($T$) is expected for CeAu$_{2}$Si$_{2}$ around this $p$.
In addition, it is worth mentioning that, after a partial depressurization from 27.8 down to $\sim$17.6 GPa, while $T^{\rm *}$ is almost unaffected, the $S$($T$) magnitude is reduced concomitantly with an increase of $\rho_{\rm 0}$,
in line with a Nordheim-Gorter type relation \cite{Note5}.
Let us note the weak $p$-dependence of the temperature $T^{\rm *}$, to which we will return below.

For $p$ $\geqslant$ 18.8 GPa, the trend in $S(T)$ versus $p$ is reminiscent of previous observations in Ce-based KLS \cite{CeCu2Si2,CeAl3,CeCu2Ge2andCePd2Si2,CeCu2,CeRu2Ge2}.
While the positive $S$($T$) keeps growing, the negative contribution decays and finally vanishes for $p$ $>$ 23.2 GPa.
In fact, $S$($T$) starts to change back to positive value at low $T$ already for $p$ $\geqslant$ 16.7 GPa.
Following the sign change, $S$($T$) increases with decreasing $T$ and then drops to zero due to the superconducting transition.
However, since such sign change may already occur below 1.3 K at lower $p$, its detailed study is beyond the scope of the present paper.
Incidentally, the anomaly in $S$($T$) associated with magnetic ordering is only observed at $p$ = 0.

At room $T$, $S$ at 27.8 GPa ($\sim$ 50$\mu$V/K) is $\sim$20 higher than at $p$ = 0.
By contrast, there is only a fivefold enhancement in $\rho$.
Together, these results signify a giant $p$-induced increase ($\sim$80 times) in the power factor $S^{2}$/$\rho$.
\begin{figure}
\includegraphics*[width=8.3cm]{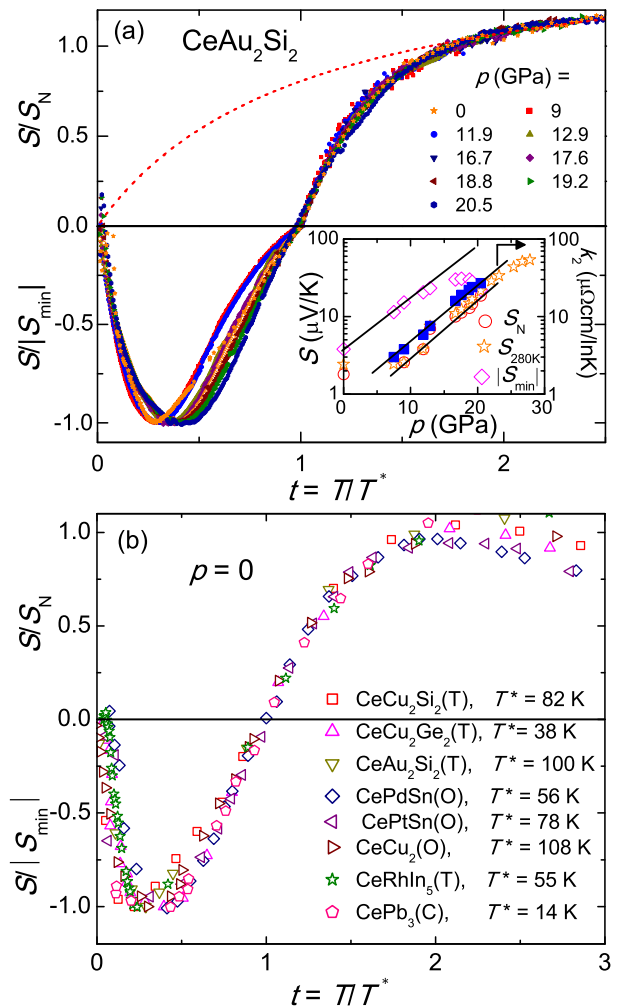}
\caption{(Color online)
(a) The normalized $S$($T$) data for $p$$ \leqslant$ 22.1 GPa as a function of $T$/$T^{\rm *}$. Data at 7.5 GPa are not considered due to possible gasket relaxation after initial pressurization.
The red dashed line denotes the fit from an empirical formula for dilute Kondo systems (see text).
The inset shows the $p$-dependence of the $S_{\rm N}$, $S_{\rm 280K}$, and $\mid$$S_{\rm min}$$\mid$ and $k_{\rm 2}$ slope above $T_{2}^{\rm max}$.
Note that both vertical axes are in logarithmic scale, and the solid lines are a guide to the eyes.
(b) the normalized $S$($T$) versus $T$/$T^{\rm *}$ for a number of Ce-based Kondo-lattice compounds at ambient pressure.
The letter in parentheses denotes the crystal symmetry, with T for tetragonal, O for orthorhombic, C for cubic.
}
\label{fig3}
\end{figure}

It is known that $S$($T$) of dilute Kondo alloys with 3$d$-impurities, when normalized by a factor of 1/$S_{\rm N}$, follows a universal function $f$($T$/$\Theta$),
where the temperature $\Theta$ characterizes the coupling between the impurity local moments and the conduction electrons \cite{Kondoalloyscaling}.
A similar situation is expected concerning 4$f$ impurities, although no experimental observations have been reported to date \cite{Zlaticbook}.
The above $\rho$($T$) results which clearly demonstrate that, CeAu$_{2}$Si$_{2}$ behaves as a Kondo alloy above the temperature $T_{i}^{\rm max}$ over a broad $p$ range,
leads us to examine whether such a scaling exists in this compound.

As shown in the main panel of Fig. 3(a), it turns out that for $T$ $\geqslant$ $T^{\rm *}$ ($S$ $\geqslant$ 0) and $p$ up to 20.5 GPa, the normalized $S$/$S_{\rm N}$ data fall on a single curve when plotted as a function of $t$ = $T$/$T^{\rm *}$.
Note that, in our case, $S_{\rm N}$ is set as $S$(1.79$T^{\rm *}$), which is the same as, or very close to, the value of $S_{\rm 280 K}$ for different $p$.
At $t$ $>$ 1.8, the scaling curve can be fitted by the empirical formula $S$ $\sim$ $T$/($T$ + $T^{\rm *}$) for dilute Kondo systems \cite{Kondoformula}, indicating that $S$($T$) is ascribed to the incoherent Kondo effect at sufficiently high $T$.
At $t$ $<$ 1, where $S(T)$ is likely governed by the thermal depopulation of the two upper CF doublets, a scaling is found when plotting $S$/$\mid$$S_{\rm min}$$\mid$ instead of $S$/$S_{\rm N}$ against $t$.
The quality of the data collapse is less good than for $t$ $>$ 1, probably due to the interference of the two contributions to the negative $S(T)$ minimum mentioned above.
For $p$ $\geqslant$ 22.1 GPa, no such scalings are observed at either low or high $T$, which we ascribe to the delocalization of Ce-4$f$ electrons \cite{RenPRX}.

The inset of Fig. 3(a) shows the resulting $S_{\rm N}$, $S_{\rm 280K}$, and $\mid$$S_{\rm min}$$\mid$ plotted as a function of $p$, together with the slope $k_{\rm 2}$ for comparison.
It is striking that $S_{\rm N}$, $S_{\rm 280K}$ and $k_{\rm 2}$ exhibit the same exponential dependence on $p$ for 9 $\leqslant$ $p$ $\leqslant$ 20.5 GPa, and so does $\mid$$S_{\rm min}$$\mid$ for $p$ $\leqslant$ 9 GPa .
According to Ref. \cite{scattering1}, $k_{\rm 2}$ $\propto$ $n^{2}(E_{\rm F})$$\mid$$J$$\mid^{3}$, where $n(E_{\rm F})$ is the density of the states at Fermi level.
Assuming a $p$-independent $n(E_{\rm F})$, we have $S_{\rm N}$ $\sim$ $S_{\rm 280K}$ $\propto$ $\mid$$S_{\rm min}$$\mid$ $\propto$ $\mid$$J$$\mid^{3}$ over a given $p$-range,
which provides strong evidence that, at both low and high $T$, the $S(T)$ magnitude enhancement is due to the increase of $J$.
In the high-$T$ limit, this is consistent with the theoretical calculation of Kondo $S(T)$ for non-interacting $f$-electron spins \cite{Fischer}.
Above 9 GPa, $\mid$$S_{\rm min}$$\mid$ deviates from the exponential behavior and tends to saturate, reflecting the competition between the positive and negative $S$($T$).
Similarly, $S_{\rm 280K}$ saturates above 26.7 GPa on approaching the value of $k_{\rm B}$/$e$.

In Fig. 3(b), we compare the scaled $S(T)$ at $p$ = 0 of various Ce-based KLS, including CeAu$_{2}$Si$_{2}$, CeCu$_{2}$ \cite{CeCu2}, CeCu$_{2}$Si$_{2}$ \cite{unpublished}, CeRnIn$_{5}$ \cite{unpublished}, CeCu$_{2}$Ge$_{2}$ \cite{CePb3TEP}, CePb$_{3}$\cite{CePb3TEP}, CePdSn \cite{CePdSnTEP}, and CePtSn \cite{CePdSnTEP}.
One can see that all data sets collapse on the same curve up to $t$ $\sim$ 2, despite the fact that $T^{\rm *}$ changes by nearly one order of magnitude.
This trend exists for CeAl$_{2}$ \cite{CeAl2TEP}, CeAl$_{3}$ \cite{CeAl3TEP}, CeCu$_{5}$Au \cite{CeCu5AuTEP} and even $\beta$-Ce \cite{betaCe}, and this list being by no means exhaustive.
These results clearly demonstrate that the scaling of $S(T)$ is widely applicable to Ce-based KLS when $J$ is sufficiently small, independently of the crystal structure or the local environment of the Ce ions.
Furthermore, as found in CeAu$_{2}$Si$_{2}$, it is expected that such a scaling still holds for these systems within a certain $p$-range.
For example, this is the case of CeCu$_{2}$Ge$_{2}$ \cite{CeCu2Ge2andCePd2Si2}.

\begin{figure}
\includegraphics*[width=8.7cm]{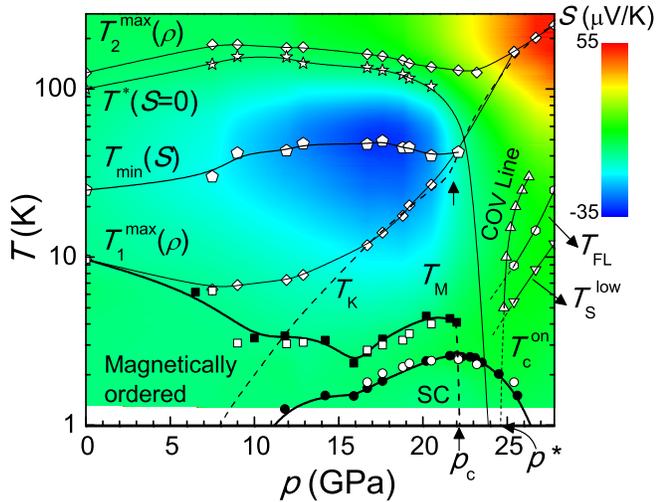}
\caption{(Color online)
$p-T$ phase diagram of CeAu$_{2}$Si$_{2}$ determined from the combination of resistivity and thermopower data.
A contour map of the thermopower data is also included.
Closed symbols represent the results from Ref. \cite{RenPRX}.
$T_{\rm K}$ is calculated from $T_{\rm K}$ $\sim$ exp[$-$$\frac{1}{NJn(E_{\rm F})}$] (see text).
The determination of $T_{\rm FL}$ and the crossover(COV) line can be found in the Supplemental Material \cite{SM}.
Note that the vertical axis is in logarithmic scale.
}
\label{fig4}
\end{figure}

To gain more insight into the $S(T)$ evolution, we constructed a comprehensive $p$-$T$ phase diagram (PD) of CeAu$_{2}$Si$_{2}$ by combining both $S$($T$) and $\rho$($T$) data, as shown in Fig. 4.
We will first discuss the driving parameter $T_{\rm K}$.
As stated previously \cite{scattering1,RenPRX}, the temperature $T_{2}^{\rm max}$ scales approximately the overall CF splitting $\Delta_{\rm CF}$,
while $T_{1}^{\rm max}$ gives an indication of $T_{\rm K}$ above 16 GPa when it rapidly grows to become much larger than $T_{\rm M}$.
With increasing $p$, the Kondo contribution to $\rho$($T$) at $T_{1}^{\rm max}$ finally dominates over that at $T_{2}^{\rm max}$, and the system enters the IV regime with a trend to recover the full degeneracy of the Ce-4$f$$^{1}$ multiplet. At lower $T$, this trend is marked by the crossover (COV) line defined from the $\rho$($p$) drop due to the 4$f$ electron delocalization \cite{RenPRX,SM,seyfarth}.
On the other hand, $T_{\rm K}$($p$) can be estimated as $T_{\rm K}$($p$) $\sim$ $D$exp[$-$$\frac{1}{NJ(p)n(E_{\rm F})}$] \cite{bandwith}, where $D$ is the bandwith, $N$ is the degeneracy of the 4$f$-state and $J$($p$) = $J_{0}$$\times$$\sqrt[3]{10^{0.073p}}$ is obtained from the fitting of $k_{\rm 2}$ data in Fig.1 (b). In order to qualitatively take into account the COV, $N$ is assumed to increase linearly with $p$ from $N$ = 2 at $p_{\rm c}$ to $N$ = 6 at 30 GPa. With the adjusted parameters $D$ = 700 K and $J_{0}$$n(E_{\rm F})$ = 0.05, the calculated $T_{\rm K}$ reproduces reasonably well the $T_{\rm 1}^{\rm max}$ for $p$ $>$ 16 GPa \cite{Note2}, confirming that $T_{\rm K}$ is increased by a factor of $\sim$40 over the huge $p$-range of SC in CeAu$_{2}$Si$_{2}$. Surprisingly, whereas no anomaly associated with the resistivity maximum at $T_{\rm 1}^{\rm max}$($T_{\rm K}$) can be observed in $S$($T$) for $p$ $<$ $p_{\rm c}$, this anomaly seems to be present in CeRu$_{2}$Ge$_{2}$ \cite{CeRu2Ge2}. In fact, the above results strongly suggest that the Kondo effect is mostly reflected in the magnitude, rather than the $T$-dependence, of $S(T)$ in CeAu$_{2}$Si$_{2}$ below $p_{\rm c}$.

For $p$ $\geqslant$ 25.4 GPa, the position of the high-$T$ maximum in $S$($T$) agrees roughly with the $T_{1}^{\rm max}$ of the resistivity maximum, and gives an independent estimation of $T_{\rm K}$. By contrast, the temperature $T_{\rm S}^{\rm low}$ of the low-$T$ maximum is only half of $T_{\rm FL}$ and $\sim$5\% of $T_{\rm K}$, where the $T$-dependent $\rho$($T$) term is still very small. Hence its origin is unlikely linked to the Kondo effect, but may be ascribed to short range magnetic correlations or to the crossover between the zero-$T$ term $S_{\rm 0}$ \cite{S01,S02} and the usual Kondo term $S_{\rm mag}$.
To better clarify this issue, the sister compounds CeCu$_{2}$$X$$_{2}$ should be re-investigated, given that such maximum appears at a much lower $p$ and hence can be studied in a more extended $p$-range \cite{CeCu2Si2,CeCu2Ge2andCePd2Si2}.

As already noted above, both $T^{\rm *}$ and $T_{\rm min}$ vary weakly with $p$ and behave very similarly to $T_{2}^{\rm max}$, unlike $T_{\rm K}$, indicating that these quantities are controlled by $\Delta_{\rm CF}$.
In this respect, the two terms contributing to the negative $S$($T$) minimum might correspond to the two excited CF levels at $\sim$190 and $\sim$250 K \cite{deltaCF}, each of them producing a minimum at $T_{\rm min}$ $\sim$ $\Delta_{\rm CF}$/6. If this is the case, this may also help to understand why our $k_{\rm 2}$/$k_{\rm 1}$ ratio is much smaller than the predicted value of 11.7 \cite{Note3}. On the other hand, the weak $p$-variation of $T^{\rm *}$  explains the difference in the observed scaling behavior between here and Ref. \cite{Kondoalloyscaling} where the CF effect is absent.

To our knowledge, there is currently no theory that can account for the observed features in $S(T)$, especially at low $T$.
The most elaborated calculations by Zlatic \emph{et al.} indicate that the sign change in $S(T)$ is a manifestation of the crossover from the weak-coupling local-moment regime to the strong-coupling Fermi-liquid regime with decreasing $T$ \cite{Zlatic2}. While this is in a qualitative agreement with the experimental results at high $T$, it is difficult to reconcile the weak $p$-dependence of $T^{\rm *}$ and $T_{\rm min}$ with the rapid growth of $T_{\rm K}$ over the broad $p$-range.
Furthermore, the shape of $S$($T$) in the crossover region cannot be determined in their calculations so that a direct comparison with the scaling behavior is not possible.
On the other hand, according to the semiphenomenological model developed by Fischer \cite{Fischer},
the dominant contribution to negative $S(T)$ stems from interacting spin pairs. However, the model predicts that $T^{\rm *}$ should scale with $T_{\rm K}$, which is at odds with our observations.
Nevertheless, since the model considers only spin interactions, it will be interesting to investigate how the situation changes when the CF effect is included.

Another salient feature illustrated in Fig. 4 is that the large negative $S(T)$ minimum (the blue region) is located right above the superconducting domain up to almost $p_{\rm c}$.
A very similar situation is found when $S(T)$/$T$ is concerned \cite{Note4}.
Actually, as suggested in Fig. 3, this appears to be a common feature of the normal state of prototypical Ce-based $p$-induced superconductors.
For CeAu$_{2}$Si$_{2}$, $\mid$$S_{\rm min}$$\mid$ and $T_{\rm c}$ exhibit a qualitatively similar $p$-dependence and hence one can speculate that this $S$($T$) minimum is intimately linked to SC.
Here it is noted that $T^{\rm *}$ vanishes at a $p$ considerably higher than $p_{\rm c}$ and shows a very different $p$-dependence from that of $T_{\rm M}$, indicating that the negative $S(T)$ is not related straightforwardly to magnetic fluctuations. In this regard, the understanding of its physical origin may help to elucidate the pairing mechanism for these materials.
Also, the scaling relation shown in Fig. 3 could serve as an empirical guide for the search of new Ce-based $p$-induced superconductors. For example, SC might be expected in Ce(Pt/Pd)Sn ($T_{\rm N}$ $\sim$ 7 K) and CePb$_{3}$ ($T_{\rm N}$ $\sim$ 1.1 K) under $p$, provided sufficiently high-quality crystals can be obtained.

Finally, it should be noted that substantial differences exist between the normal-state properties of CeAu$_{2}$Si$_{2}$ and CeCu$_{2}$$X_{2}$ despite their similarities \cite{RenPRX}. In particular, regardless its smaller overall $\Delta_{\rm CF}$, $T^{\rm *}$ of CeAu$_{2}$Si$_{2}$ ($\sim$120-150 K) is almost twice that of CeCu$_{2}$$X_{2}$ ($\sim$40-80 K). Moreover, the low-$\rho_{\rm 0}$ of CeAu$_{2}$Si$_{2}$ in comparison with CeCu$_{2}$Ge$_{2}$ suggests a longer
mean free path, which is more favorable for unconventional SC \cite{RenPRB}. Given that the interaction leading to the negative $S(T)$ may also be involved in the superconducting pairing, these factors could be the key to understanding the exotic ground-state properties of CeAu$_{2}$Si$_{2}$ under $p$.

\section{IV. Conclusion}
In summary, we have studied systematically the high-$p$ thermopower and resistivity of CeAu$_{2}$Si$_{2}$ up to 27.8 GPa.
For the first time, a scaling behavior is found in the $S(T)$ data below 20.5 GPa as a function of a reduced temperature $T$/$T^{\rm *}$.
The comparison with $\rho(T)$ results shows that the $S(T)$ magnitude is determined by the Kondo coupling $J$, while the CF splitting $\Delta_{\rm CF}$ controls the characteristic temperatures of the sign change ($T^{\rm *}$) and the negative minimum ($T_{\rm min}$).
Up to almost $p_{\rm c}$, a large negative $S(T)$ minimum regularly precedes the superconducting transition, suggesting that the two phenomena are closely related.
Furthermore, the scaling relation is shown to hold up to 2$T^{\rm *}$ for $S$($T$) at ambient $p$ of related systems with diverse crystal structures, testifying to its wide applicability.
Our work demonstrates that thermopower can be useful in probing the $p$-evolution of CF energy scale in Ce-based KLS, but should be used with caution in estimating the $T_{\rm K}$ of these materials.
This calls for new theoretical understandings.

\section{ACKNOWLEDGEMENT}
\begin{acknowledgments}
We acknowledge enlightening discussions with V. Zlatic and R. Monnier, as well as financial support from the Swiss National Science Foundation through Grant No. 200020-137519.
\end{acknowledgments}

\end{document}